\documentclass[11pt]{article}
\usepackage[english]{babel}
\usepackage{amstext,amsfonts,amsmath,amssymb}

\newfont{\gl}{eufm10 scaled 1200} %% gothic fonts

\numberwithin{equation}{section}
\providecommand{\bysame}{\leavevmode\hbox to3em{\hrulefill}\thinspace}
\providecommand{\MR}{\relax\ifhmode\unskip\space\fi MR }
\providecommand{\href}[2]{#2}

\newcommand{\dd}{{\rm d}}

\begin{document}

\title{{\bf Star products and branes in Poisson-Sigma models}}

\author{Iv\'an Calvo\thanks{Research supported by grant FPU, MEC
(Spain).} \ and Fernando Falceto\thanks{Research supported by grant
FPA2003-02948, MEC (Spain).}\\{}\\Departamento de F\'{\i}sica Te\'orica\\
Universidad de Zaragoza\\ E-50009 Zaragoza, Spain \\
\\E-mail: {\tt icalvo@unizar.es,
falceto@unizar.es}
}

\date{}

\maketitle

\begin{abstract}
We prove that non-coisotropic branes in the Poisson-Sigma model are
allowed at the quantum level. When the brane is defined by
second-class constraints, the perturbative quantization of the model
yields Kontsevich's star product associated to the Dirac bracket
on the brane. Finally, we present the quantization for a general
brane.
\end{abstract}

\section{Introduction}
In their celebrated paper \cite{CatFel99} Cattaneo and Felder gave a
field theoretical interpretation of Kontsevich's formula (\cite{Kon})
for the deformation quantization of a Poisson manifold $(M,\Pi)$ where
$\Pi$ stands for the Poisson structure. The field theory derivation
involves the so-called Poisson-Sigma model
(\cite{Ike},\cite{SchStr94}). This is a two-dimensional topological
field theory defined on a surface $\Sigma$ whose target is a Poisson
manifold $(M,\Pi)$. The field content is a bundle map from $T\Sigma$
to $T^*M$. Cattaneo and Felder showed that Kontsevich's formula can be
obtained from Feynman expansion of certain Green's functions when
$\Sigma$ is the unit disc $D$ and the base map $X:\Sigma \rightarrow
M$ has free boundary conditions.

The same authors proved in \cite{CatFel03} that the
non-symmetry-breaking boundary conditions of the Poisson-Sigma model
are given by coisotropic branes, i.e. submanifolds defined by {\it
first-class constraints}. In this case the quantization of the model
is related to the deformation quantization of the coisotropic
submanifold.

We proved recently (\cite{CalFal04}) that classically the field $X$
can be consistently restricted at the boundary $\partial\Sigma$ to an
almost arbitrary submanifold $C$. It turns out that the symplectic
structure on the reduced phase space of the model is related to the
Poisson bracket canonically induced on (a subset of) $C^\infty(C)$.

On the light of these results it is natural to conjecture that the
perturbative quantization of the model with general $C$ be related to
the deformation quantization of the induced Poisson bracket on
(certain functions on) $C$. In this paper we study in detail the case
in which $C$ can be defined by a set of {\it second-class constraints}
which is, in some sense, opposite to the coisotropic one. The
quantization of the coisotropic case (\cite{CatFel03}) presents some
intricacies due to the fact that gauge transformations do not vanish
at the boundary. If $C$ is defined by second-class constraints ({\it
second-class brane}) they do vanish and one would expect to have a
clean quantization recovering Kontsevich's formula, this time not for
$\Pi$ but for the Dirac bracket on $C$. We show that this expected
result holds and that it emerges in quite a different way from the
coisotropic case. Finally, we give the quantization of the
Poisson-Sigma model with a general brane defined by a mixture of first
and second class constraints.

The paper is organized as follows:

In Section 2 we give a brief introduction to Poisson geometry and
Poisson reduction, as well as to the problem of deformation
quantization and Kontsevich's solution.

Section 3 recalls the results of \cite{CalFal04} on the study of the
classically consistent boundary conditions for the Poisson-Sigma model.

Section 4 is devoted to the problem of perturbative quantization of
the Poisson-Sigma model with non-coisotropic branes. First, we prove
that in the perturbative expansion as defined in \cite{CatFel99} and
\cite{CatFel03} the propagator does not exist when second-class
constraints are involved. We show that with slight modifications a
well-defined perturbative expansion can be given. However, this first
solution is still too naive. In subsection 4.2.2 we take a more
original approach and show that the perturbative quantization of the
model with target $(M,\Pi)$ and with a second-class brane gives
Kontsevich's formula corresponding to the Dirac bracket on $C$
obtained by reduction from $\Pi$.

\section{Reduction of Poisson manifolds} \label{PoissonGeometry}

This subsection is a brief summary of some results on Poisson
reduction presented in \cite{CalFal04}. We refer the reader to that
paper for details.

Given a Poisson manifold $(M,\Pi)$ and a closed submanifold
$C\hookrightarrow M$, we would like to know whether $\Pi$ defines in a
canonical way a Poisson bracket on $C^\infty(C)$ or at least on a
subset of it.

We adopt the notation ${\cal A} = C^{\infty}(M)$ and take the ideal 
(with respect to the point-wise product of functions in ${\cal A}$)
$${\cal I} =\{f \in {\cal A} \vert f(p) = 0,\ p \in C\}$$

We view $C^{\infty}(C)$ as ${\cal A}/{\cal I}$.

Define ${\cal F}\subset{\cal A}$ as the set of {\it first-class
functions}, also called the {\it normalizer} of ${\cal I}$,
$${\cal F}=\{f\in{\cal A} \vert \{f,{\cal I}\}\subset{\cal I}\}.$$

Note that due to the Jacobi identity and the Leibniz rule ${\cal F}$
is a Poisson subalgebra of ${\cal A}$ and ${\cal F}\cap{\cal I}$ is a
Poisson ideal of ${\cal F}$. Then, we have canonically defined a
Poisson bracket in the quotient ${\cal F}/({\cal F}\cap{\cal I})$.

Now, we define the map
\begin{eqnarray}\label{mapphi}
 \begin{matrix}\phi:&{\cal F}/({\cal F}\cap{\cal I})
&\longrightarrow&{\cal A}/{\cal I}\cr
&f+{\cal F}\cap{\cal I}&\longmapsto&f+{\cal I}
\end{matrix}
\end{eqnarray}
which is an injective homomorphism of abelian, associative algebras with unit 
and then induces a Poisson algebra structure $\{.,.\}_C$ on the 
image, i.e.:
\begin{eqnarray}\label{ourDirac}
\{ f_1+{\cal I}, f_2+{\cal I}\}_{_C}=
\{ f_1, f_2\}+{\cal I}.\qquad f_1,f_2\in {\cal F}.
\end{eqnarray}

\vspace{0.3 cm}

{\it Remark:} Note that the elements of ${\cal F}\cap{\cal I}$ are, in
the language of physicists, the generators of {\it gauge
transformations} or, in Dirac's terminology, the {\it first-class
constraints}.

\vspace{0.3 cm}

In case $\phi$ is onto we have endowed $C$ with a canonical Poisson
structure. As shown in ref. \cite{CalFal04} this situation is
equivalent to the existence of what Vaisman defines as an {\it
algebraically $\Pi$-compatible normal bundle} of $C$, see ref.
\cite{Vai02} for details. A different strategy for endowing $C$ with a
canonical Poisson structure is via the reduction of Dirac structures,
see ref. \cite{CraFer}. Here is denoted {\it Poisson-Dirac} the
submanifold for which this reduction actually defines a Poisson
structure on $C$.  It is easy to see that if map $\phi$ is onto, $C$
is Poisson-Dirac and the induced Poisson structure obtained in both
ways is the same. We refer the reader to ref. \cite{CalFal04} for more
details on the relations between the two approaches.

In general, however, $\phi$ is not onto and $C$ cannot be made a
Poisson manifold. What we have is a Poisson bracket on $\phi({\cal
F}/({\cal F}\cap{\cal I}))\subseteq {\cal A}/{\cal I}$. The image of
$\phi$ is not easy to characterize in the general case but, as we
shall see next, it has a nice interpretation if certain regularity
conditions are met.

Let $N^*C$ (or ${\rm Ann}(TC)$) be the {\it conormal bundle} of $C$
(or {\it annihilator} of $TC$), i.e. the subbundle of the pull-back
$i^{*}(T^{*}M)$ consisting of covectors that kill all vectors in
$TC$. Define also the set of {\it gauge-invariant functions}
$${\cal A}_{inv}:=\{ f\in{\cal A} | \{f,{\cal F}\cap{\cal I}\}
\subset{\cal I}\}.$$

We have the following

\vskip 0.2 cm

{\bf Theorem:}

\vskip 0.2 cm

{\it \parindent 0pt If ${\rm dim}(\Pi^\sharp_p(N_p^*C)+ T_pC)$ is
constant for every $p\in C$, then $\phi ({{\cal F}}/{{\cal
F}}\cap{{\cal I}})={\cal A}_{inv}/{\cal I}$. In other words, the image
of $\phi$ are the gauge-invariant functions restricted to $C$.}

\vskip 0.3 cm

{\it Proof:} See \cite{CalFal04}.
$\,$\hfill$\Box$\break

\vspace{0.5 cm}

The meaning of what we shall call the {\it strong regularity
condition}
\begin{eqnarray} \label{regularity}
{\rm dim}(\Pi^\sharp_p(N_p^*C)+ T_pC)=k+\dim(C),\quad \forall p\in C
\end{eqnarray}
for a non-negative constant $k$, is clarified by noticing that it
allows to choose in a neighborhood ${\cal U}\subset M$ of every $p\in
C$ adapted local coordinates on $M$, $(X^a,X^\mu,X^{A})$, with
$a=1,\dots,\dim(C)$, $\mu=\dim(C)+1,\dots,\dim(M)-k$ and
$A=\dim(M)-k+1,\dots,\dim(M)$, verifying:

\vskip 0.2cm

(i)  $C\cap{\cal U}$ is defined by $X^\mu=X^{A}=0$.

\vskip 0.2cm

(ii) $\{X^{\mu},X^{\nu}\}\vert_{C\cap{\cal
U}}=\{X^{\mu},X^A\}\vert_{C\cap{\cal U}}=0$,
i.e. $X^{\mu}$ are {\it first-class constraints}.

\vskip 0.2cm

(iii) ${\rm det}(\{X^A,X^B\}(p))
\neq 0,\quad \forall p\in{C\cap{\cal U}}$, 
i.e. $X^A$ are {\it second-class constraints}.

\vskip 0.2cm

It is clear that in these adapted coordinates the Poisson structure
satisfies:
\begin{eqnarray} \label{localcoor}
\Pi^{\mu\nu}\vert_{{C}\cap{\cal U}} = 0,\quad
\Pi^{A\mu}\vert_{{C}\cap{\cal U}} = 0,\quad {\rm
det}(\Pi^{AB})\vert_{{C}\cap{\cal U}} \neq 0
\end{eqnarray}

\vskip 0.2 cm

Notice also (Lemma 1 of ref. \cite{CalFal04}) that the strong regularity 
condition is equivalent to 
\begin{eqnarray}\label{equivalence}
N_p^*C \cap \Pi_p^{\sharp
-1}(T_pC)=\{(\dd f)_p | f\in{\cal F}\cap{\cal I}\},\quad \forall p\in C
\end{eqnarray}

{\it Remark:} $C$ is said to be coisotropic if
$\Pi^\sharp(N^*C)\subseteq TC$. For such $C$ the strong regularity
condition (\ref{regularity}) is obviously satisfied and every
constraint is first-class. In addition, (\ref{ourDirac}) is the
original bracket on $M$ restricted to the gauge-invariant functions on
$C$. The case of free boundary conditions, $C=M$, is an extreme
example of coisotropic submanifold.

\vskip 0.2 cm

Later on we shall be concerned with the situation in which every
constraint defining $C$ is second-class, i.e. there are no Greek
indices and the strong regularity condition is fulfilled.  We call
such $C$ a {\it second-class submanifold} or {\it second-class
brane}. In this case the matrix of the Poisson brackets of the
constraints $\Pi^{AB}=\{X^A,X^B\}$ is invertible on $C$. Defining on
$C$ the matrix $\omega_{AB}$ by $\omega_{AB}\Pi^{BC}=\delta_A^C$ the
Poisson bracket (\ref{ourDirac}) can be written locally:
\begin{eqnarray}\label{Dirac}
\{f +{\cal I},f'+{\cal I}\}_{_C}=\{f,f'\}-\{f,X^A\}\omega_{AB}
\{X^B,f'\}+{\cal I}
\end{eqnarray}
which is the usual definition of the {\it Dirac bracket} restricted to
$C$.  In this case every function on $M$ is trivially gauge-invariant
(since ${\cal F}\cap{\cal I}=0$), the image of $\phi$ is $C^\infty(C)$
and we get a Poisson structure on $C$. In adapted coordinates the
components of the canonical 
Poisson tensor on $C$ corresponding to (\ref{Dirac})
are given by:
\begin{eqnarray} \label{DiracComponents}
\Pi^{ab}_{\cal D}=\Pi^{ab}-\Pi^{aA}\omega_{AB}\Pi^{Bb}
\end{eqnarray}
where the subscript $\cal D$ stands for Dirac.

\vskip 0.2 cm

When first-class constraints are present one can still use formula
(\ref{DiracComponents}). Given a choice of adapted coordinates $\{X^a,
X^\mu, X^A\}$ the expression
\begin{eqnarray} \label{DiracCompFirst}
\Pi^{pq}_{\cal D}=\Pi^{pq}-\Pi^{pA}\omega_{AB}\Pi^{Bq}
\end{eqnarray}
with the indices $p, q=1,\dots,\dim(M)-k$ running over $a$ and $\mu$
values, defines a Poisson bracket in the submanifold $C'$ on which the
second-class constraints vanish (we assume $\det(\Pi^{AB})\not=0$ on
$C'$).  The submanifold $C'$ is not uniquely defined, as it depends of
the concrete choice of the set of second class constraints.  $C$ is
now a coisotropic submanifold of $C'$ and the Poisson algebra induced
by $\Pi_{\cal D}$ on the gauge invariant functions on $C$ is indeed
canonical, independent of the choice of $C'$, and equals the one given
by (\ref{ourDirac}).

One can also extend the Poisson tensor to a tubular neighborhood of
$C'$ by taking $\Pi_{\cal D}^{Ap}=\Pi_{\cal D}^{AB}=0$.  If one
considers the tubular neighborhood equipped with the Dirac bracket,
$C$ is coisotropic in it and $C'$ is a Poisson submanifold.

\vskip 0.2 cm

For later purposes it is useful to consider the following {\it weak
regularity condition}:
\begin{eqnarray} \label{weakregularity}
\dim\{(\dd f)_p | f\in{\cal F}\cap{\cal I}\}=k+\dim(M)-\dim(C),\quad \forall
p\in C
\end{eqnarray}
for some non-negative constant $k$.

That the strong regularity condition (\ref{regularity}) implies the
weak one with the same value for the constant $k$ is clear from
(\ref{equivalence}).

The weak regularity condition is equivalent to the existence of local
coordinates on a tubular neighborhood of every patch of $C$ with a
maximal (and constant) number of coordinates which are
first-class constraints. In other
words, (\ref{weakregularity}) holds if and only if there exist local
coordinates satisfying (i), (ii) as above and

\vskip .2cm
(iii)$'$ ${\rm det}(\{X^A,X^B\}(p)) \neq 0\quad \text{for
} p \text{ in an open dense subset of } {C\cap{\cal U}}$.
\vskip .2cm

However, in general the weak regularity condition is not enough to
guarantee that $\phi({\cal F}/{{\cal F}\cap{\cal I}})={\cal
A}_{inv}/{\cal I}$.

\section{Poisson-Sigma models on surfaces with boundary} \label{PSmodels}
 
The Poisson-Sigma model is a two-dimensional topological Sigma model
defined on a surface $\Sigma$ and with a finite dimensional Poisson
manifold $(M,\Pi)$ as target. The fields of the model are given by
a bundle map $(X,\eta): T\Sigma \rightarrow T^{*}M$ consisting of a
base map $X:\Sigma \rightarrow M$ and a 1-form $\eta$ on $\Sigma$ with
values in the pullback by $X$ of the cotangent bundle of $M$.  The
action functional has the form\footnote{We adopt throughout this
paper the notation and sign conventions of Cattaneo and Felder's paper
\cite{CatFel99}.}
\begin{eqnarray} \label{PS}
S(X,\eta)=\int_\Sigma \langle \eta,\wedge \dd X \rangle +
{1\over2}\langle\Pi\circ X,\eta\wedge\eta\rangle,
\end{eqnarray}
where $\langle\cdot,\cdot\rangle$ denotes the pairing between vectors and
covectors of $M$.

If $X^{i}$ are local coordinates in $M$, $\sigma
^{\kappa},\ \kappa=1,2$ local coordinates in $\Sigma$, $\Pi^{ij}$ the
components of the Poisson structure in these coordinates and
$\eta_{i}=\eta_{i\kappa}d{\sigma}^{\kappa}$, the action reads
\begin{eqnarray} \label{PScoor}
S(X,\eta)=\int_\Sigma  \eta_{i}\wedge \dd X^{i}+
{1\over2}\Pi^{ij}(X)\eta_{i}\wedge \eta_{j}
\end{eqnarray}

The equations of motion in the bulk are:
\begin{subequations}\label{eom}
\begin{align}
&\dd X^{i}+\Pi^{ij}(X)\eta_{j}=0 \label{eoma} \\ 
&\dd\eta_{i}+{1\over2}\partial_{i}\Pi^{jk}(X)\eta_{j}
\wedge \eta_{k}=0 \label{eomb}
\end{align}
\end{subequations}

The infinitesimal transformations
\begin{subequations}\label{symmetry}
\begin{align}
&\delta_{\epsilon}X^{i}=
\Pi^{ij}(X)\epsilon_{j}\label{symmetrya}\\
&\delta_{\epsilon}\eta_{i}=-\dd \epsilon_{i}-\partial_{i}\Pi^{jk}(X)\eta_{j}
\epsilon_{k}\label{symmetryb}
\end{align}
\end{subequations}
where $\epsilon=\epsilon_{i}dX^i$ is a section of $X^*(T^*M)$,
change the action (\ref{PScoor}) by a boundary term
\begin{eqnarray} \label{symmS}
\delta_{\epsilon}S = -\int_\Sigma \dd (\dd X^i \epsilon_i).
\end{eqnarray}

Notice that
\begin{subequations}\label{OpenAlgebra}
\begin{align}
 &[\delta_\epsilon,\delta_{\epsilon'}]X^i=\delta_{[\epsilon,\epsilon']^*}
X^i \label{commga} \\
&[\delta_\epsilon,\delta_{\epsilon'}]\eta_i=
\delta_{[\epsilon,\epsilon']^*} \eta_i
-\epsilon_k\epsilon_{l}'\partial_i\partial_j
\Pi^{kl}(\dd X^{j}+\Pi^{js}(X)\eta_{s})\label{commgb}
\end{align}
\end{subequations}
where $[\epsilon,\epsilon']^{*}_k := -\partial_k\Pi^{ij}(X)
\epsilon_i\epsilon'_j$.  The term in parenthesis in \eqref{commgb} is
the equation of motion \eqref{eoma}. Hence, the commutator of two
transformations of type (\ref{symmetry}) is a transformation of the
same type only on-shell and the gauge transformations
(\ref{symmetry}) form an open-algebra.

If $\Sigma$ has a boundary a new term appears in the variation of the
action under a change of $X$ when performing the integration by parts:
\begin{eqnarray} \label{boundaryterm}
\delta_{X} S=-\int_{\partial\Sigma}\delta X^i\eta_{i}+
\int_\Sigma\delta X^i(\dd\eta_{i}+
{1\over2}\partial_{i}\Pi^{jk}(X)\eta_{j}\wedge \eta_{k})
\end{eqnarray}

Let us restrict the field $X$ at the boundary to a closed submanifold
$C$ of $M$:
\begin{eqnarray}\label{BCX}
X|_{\partial \Sigma}:\partial \Sigma\rightarrow C\subset M
\end{eqnarray}

The conditions for $\eta$ should make the boundary term in
(\ref{boundaryterm}) vanish and make the equations of motion
(\ref{eoma}) consistent at the boundary.  This is achieved if we take
the following boundary conditions (BC) for the fields:

\vskip 0.2cm

(XBC)\qquad $X(m) \in C,\forall m\in \partial\Sigma$

\vskip 0.2cm

($\eta$BC)\qquad $\eta_{\bf t}(m)\in\{(\dd f)_m | f\in{\cal F}\cap{\cal I}\}
\ \forall m\in \partial\Sigma$

\vskip 0.2cm

\noindent where $\eta_{\bf t}=\eta_{i{\bf t}}\dd X^i$ is the
contraction of $\eta$ with vector fields tangent to the boundary. In
order to have gauge transformations compatible at the boundary with
the BC one must have

\vskip 0.2cm

($\epsilon$BC)\qquad $\epsilon(m)\in\{(\dd f)_m | f\in{\cal F}\cap{\cal I}\}
\ \forall m\in \partial\Sigma$

\vskip 0.2cm

Clearly, (XBC) is preserved by (\ref{symmetrya}). In
reference \cite{CalFal04} it is proven that if the strong regularity
condition (\ref{regularity}) holds, and in the adapted coordinates
described in section 2.1, the gauge transformation (\ref{symmetryb})
also preserves ($\eta$BC). Here we redo the proof assuming only the
{\bf weak regularity condition}.

We take the adapted coordinates of section 2.1 $\{X^a,X^\mu,X^A\}$
where we use $a,b,...$ for the coordinates on $C$, $\mu,\nu,...$ for
the first-class constraints and $A,B,...$ for the second-class
ones. In these coordinates $\{\dd f \vert f\in {\cal F}\cap{\cal I}\}$
is spanned locally by $\dd X^\mu$ and the BC read:

\vskip 0.2cm

(XBC)\qquad $X^\mu(m)=X^{A}(m)=0,\ \forall m\in \partial\Sigma$

\vskip 0.2cm

($\eta$BC)\qquad $\eta_{a{\bf t}}(m) = \eta_{A{\bf t}}(m)=0,\ \forall m\in
\partial\Sigma$

\vskip 0.5cm

($\epsilon$BC)\qquad $\epsilon_{a}(m) = \epsilon_{A}(m)=0,\ \forall m\in
\partial\Sigma$.

\vskip 0.5cm

The consistency of these BC require that gauge variations of
$\eta_{a{\bf t}}$ and $\eta_{A{\bf t}}$ vanish at the boundary. For first-class components
$$\delta_{\epsilon}\eta_{a{\bf t}}(m)=-\partial_{a}
\Pi^{\mu\nu}(X(m))\eta_{\mu{\bf t}}(m)\epsilon_\nu(m)=0,
\quad\forall m\in\partial\Sigma$$
because $\partial_a$ is a derivative in a direction tangent to $C$ and
$\Pi^{\mu\nu}$ vanishes on $C$.

In order to prove the vanishing at the boundary of the transformation of {\it
second-class} components
\begin{eqnarray}\label{conseta}\
\delta_{\epsilon}\eta_{A{\bf t}}(m)=-\partial_{A}
\Pi^{\mu\nu}(X(m))\eta_{\mu{\bf t}}(m)\epsilon_\nu(m)=0,\quad m\in\partial\Sigma
\end{eqnarray}
one needs the following Jacobi identity

\begin{eqnarray}
\Pi^{AB}
\partial_A\Pi^{\mu\nu}
+
\Pi^{\gamma B}
\partial_\gamma\Pi^{\mu\nu}
+
\Pi^{aB}
\partial_a\Pi^{\mu\nu}
&&\cr
+ \Pi^{A\nu}
\partial_A\Pi^{B\mu}
+
\Pi^{\gamma\nu}
\partial_\gamma\Pi^{B\mu}
+
\Pi^{a\nu}
\partial_a\Pi^{B\mu}
&&\cr
+ \Pi^{A\mu}
\partial_A\Pi^{\nu B}
+
\Pi^{\gamma\mu}
\partial_\gamma\Pi^{\nu B}
+
\Pi^{a \mu}
\partial_a\Pi^{\nu B}
&=& 0
\end{eqnarray}

Evaluating the previous expression on $C$ and using $\Pi^{\mu\nu}\vert_C =
\Pi^{\mu A}\vert_C=0$ one has
$$\Pi^{AB} \partial_A\Pi^{\mu\nu}\vert_C=0.$$

{}From the fact that $\det(\Pi^{AB})\neq 0$ in an open dense subset of
$C$ as implied by the weak regularity condition, an argument of
continuity shows that $\partial_A\Pi^{\mu\nu}\vert_C=0$ and
$\delta_{\epsilon}\eta_{A{\bf t}}$ vanishes at the boundary of $\Sigma$.

We shall call {\it weakly regular brane} to a submanifold $C$ which
satisfies the weak regularity condition. As shown above, weakly
regular branes lead to consistent boundary conditions for the Poisson
Sigma model at the classical level.

\vskip 0.5cm

\section{Quantization of the Poisson-Sigma model}

\subsection{Batalin-Vilkovisky procedure}

Here we recall the steps followed in \cite{CatFel99} to construct the
partition function of the Poisson-Sigma model\footnote{The path
integral quantization of the Poisson-Sigma model in the particular
case of 2D gravity was first carried out in \cite{KumLieVas}.}.

Consider the space of fields\footnote{In this subsection we are not
concerned with the boundary conditions of the additional fields
entering the formalism. They are discussed in detail in the next
subsection.} with a $\mathbb Z$ gradation corresponding to the ghost
number and a ${\mathbb Z}_2$ gradation corresponding to the Grassmann
parity. The standard BRST formalism for the quantization of a theory
with gauge symmetries introduces anticommuting scalar fields $\beta_i$
(ghosts) and $\gamma^i$ (antighosts) along with commuting scalar
fields $\lambda^i$ (Lagrange multipliers). The basic ghost number
assignments are:
$\text{gh}(X^i)=\text{gh}(\eta_i)=\text{gh}(\lambda^i)=0$,
$\text{gh}(\beta_i)=1$, $\text{gh}(\gamma^i)=-1$.

Now one defines an odd derivation of ghost number one $\delta_0$:
\begin{subequations}
\begin{align}
&\delta_{0}X^{i}=
\Pi^{ij}(X)\beta_{j}\\
&\delta_{0}\eta_{i}=-\dd \beta_{i}-\partial_{i}\Pi^{jk}(X)\eta_{j}
\beta_{k}\\
&\delta_0\beta_i=\frac{1}{2}\partial_i\Pi^{jk}(X)\beta_j\beta_k\\
&\delta_0\gamma^i=\lambda^i\\
&\delta_0\lambda^i=0
\end{align}
\end{subequations}
extended to functions of the fields through the Leibniz rule. The
problem is that since the gauge transformations (\ref{symmetry}) close
only on-shell, $\delta_0^2$ vanishes only on-shell and we do not have a
well-defined cohomology on the space of fields.

The extension of the BRST scheme which works for open algebras is
known as Batalin-Vilkovisky (BV) procedure\footnote{See the original
papers \cite{BatVil81},\cite{BatVil83} and the excellent exposition
included in \cite{Jon}.}. Firstly, we double the number of fields by
introducing an antifield $\varphi^+_i$ for each field $\varphi^i$
($\varphi$ stands here for X,$\eta$,$\beta$,$\gamma$ and $\lambda$)
such that $\varphi^+_i$ has Grassmann parity opposite to that of $\varphi^i$
and $\text{gh}(\varphi^+_i)=-1-\text{gh}(\varphi^i)$.

The partition function is given by
\begin{eqnarray} \label{BVZ}
Z=\int e^{\frac{i}{\hbar}S_{BV}}\delta\left(\varphi_i^+ -
\frac{\overrightarrow{\delta}\Psi}{\delta
\varphi^i}\right)\cal{D}\varphi\cal{D}\varphi^+
\end{eqnarray}
where 
\begin{eqnarray} \label{SBV}
S_{BV}&=&\int_D \eta_i\wedge\dd X^i + \frac{1}{2}\Pi^{ij}(X)\eta_i\wedge\eta_j+
X_i^+\Pi^{ij}(X)\beta_j-\\
&&\ \ -\eta^{+i}\wedge(\dd \beta_i+\partial_i\Pi^{kl}(X)
\eta_k\beta_l)-\frac{1}{2}\beta^{+i}\partial_i\Pi^{jk}(X)\beta_j\beta_k-
\notag\\&&\ \ -\frac{1}{4}
\eta^{+i}\wedge\eta^{+j}\partial_i\partial_j\Pi^{kl}(X)\beta_k\beta_l-
\lambda^i\gamma^+_i \notag
\end{eqnarray}
and the {\it gauge-fixing fermion} $\Psi$ is an anticommuting
functional of the fields of ghost number $-1$ which makes the path
integral well-defined. The canonical choice is to take $\Psi$ as the
scalar product of the antighosts and the gauge-fixing conditions.

\subsection{Quantization on the disc}

Take $\Sigma$ the unit disc $D=\{\sigma\in \mathbb{R}^2, |\sigma| \leq
1\}$. Cattaneo and Felder showed in \cite{CatFel99} that the
perturbative expansion of certain Green's functions of the
Poisson-Sigma model defined on $D$ with $C=M$ (free BC) yields
Kontsevich's $\star$-product corresponding to the Poisson manifold
$M$, namely
\begin{eqnarray}\label{GreenFunctions}
\left<f(X(0))g(X(1))\delta(X(\infty)-x)\right>=f{\star}g(x)
\end{eqnarray}
where $0$, $1$ and $\infty$ are three cyclically ordered points at the
boundary of $D$ and the expectation value is calculated using
(\ref{BVZ}). In a later work (\cite{CatFel03}) the same authors
studied the quantization with a general coisotropic brane $C$, which
turned out to be related with the deformation quantization of the
submanifold $C \hookrightarrow M$.

In both papers \cite{CatFel99},\cite{CatFel03} the Green's functions
(\ref{GreenFunctions}) are worked out in the same fashion: first, one
takes the {\it Lorentz gauge} $\dd{*\eta}=0$, where the Hodge operator
acting on 1-forms requires the introduction of a complex structure on
$D$. The Feynman expansion in powers of $\hbar$ is then performed
around the constant classical solution where $X(\sigma)=x\in C$ and
the rest of the fields vanish. In this case expanding in powers of
$\hbar$ amounts to expanding in powers of $\Pi$ or, equivalently, to
expanding around zero Poisson structure.

In the next subsection we shall try to work out (\ref{GreenFunctions})
following the steps enumerated in the last paragraph when $C$ is
non-coisotropic. We shall see that when second-class constraints are
present the propagator does not exist, but a natural redefinition of
the unperturbed (or quadratic) part of the action will yield a
well-defined perturbative expansion, showing that the non-coisotropic
branes also make sense at the quantum level. However, this leads to a
messy expression whose interpretation is far from clear. In the
subsection 4.2.2 we shall see that a change of the gauge fixing is
illuminating in order to unravel the relation between the quantization
of the Poisson-Sigma model with a non-coisotropic brane and
Kontsevich's formula.

\subsubsection{The perturbation expansion in the non-coisotropic case}

Let us take the {\it Lorentz gauge} $\dd{*\eta}=0$ as said above. The
gauge fixing fermion is then:
$$\Psi= \int_D \gamma^i\dd{*\eta_i}$$
and the gauge fixed action with the antifields integrated out is
\begin{eqnarray}\label{gfaccion}
S_{gf}&=&\int_D \eta_i\wedge\dd X^i + \frac{1}{2}\Pi^{ij}(X)\eta_i\wedge\eta_j
- *{\dd\gamma}^i
\wedge(\dd \beta_i+\partial_i\Pi^{kl}(X)
\eta_k\beta_l)- \cr
&-&\frac{1}{4}
*{\dd\gamma}^i\wedge
*{\dd\gamma}^j\wedge
\partial_i\partial_j\Pi^{kl}(X)\beta_k\beta_l-
\lambda^i\dd{*\eta}_i
\end{eqnarray}

Now write $X^i(\sigma)=x^i+\xi^i(\sigma)$ and choose
\begin{eqnarray}\label{unpgfaccion}
S_0&=&\int_D \eta_i\wedge\dd{\xi^i} 
- *\dd\gamma^i 
\wedge\dd \beta_i
-\dd{*\eta_i}\lambda^i
\end{eqnarray}
as the quadratic part which defines the propagators whereas
$S_{pert}=S_{gf}-S_0$ yields the vertices of the perturbative
expansion (see \cite{CatFel99} for
explicit expressions). Our aim is to show 
that for non-coisotropic $C$
the propagator cannot fulfill the appropriate BC.

In the adapted coordinates of section 2.1 index $i$ splits into
$a,\mu,A$ where $\xi^a$ are coordinates along the brane (free at the
boundary) and $\xi^\mu$ and $\xi^A$ are respectively first-class and
second-class coordinates transversal to the brane and must vanish at
the boundary.  For the rest of the fields we have Dirichlet boundary
conditions for $\lambda^a$, $\lambda^A$, $\eta_{a{\bf t}}$,
$(*\eta_{\mu})_{\bf t}$, $\eta_{A{\bf t}}$, $\beta_a$, $\beta_A$,
$\gamma^a$ and $\gamma^A$ and Neumann boundary conditions for
$\beta_\mu$, $\gamma^\mu$ and $\lambda^\mu$.

It is convenient to map conformally the disc onto the upper complex
half plane $H_+$ (recall the conformal invariance of $S_{gf}$) and use
a complex coordinate $z\in H_+$.  The propagators for the $\beta_i$
and $\gamma^i$ fields are given by the Green's function of the Laplacian
with Dirichlet boundary conditions for components along the brane
and for second-class constraints and Neumann boundary conditions
for first-class constraints, i.e.
\begin{eqnarray}\label{betagamma}
\langle \gamma^a(w,\bar w)\beta_b(z,\bar z)\rangle_0&=&
\frac{i\hbar}{2\pi}\delta_b^a\log\frac{|w-z|}{|w-\bar z|}\cr
\langle \gamma^\mu(w,\bar w)\beta_\nu(z,\bar z)\rangle_0&=&
\frac{i\hbar}{2\pi}\delta_\nu^\mu\log(|w-z||w-{\bar z}|)\cr
\langle \gamma^A(w,\bar w)\beta_B(z,\bar z)\rangle_0&=&
\frac{i\hbar}{2\pi}\delta_B^A\log\frac{|w-z|}{|w-\bar z|}
\end{eqnarray}

The other non vanishing components of the propagator
are conveniently expressed in terms of
the complex fields
$\zeta^j=\xi^j+i\lambda^j$ 
and $\bar\zeta^j=\xi^j-i\lambda^j$. 

The general solution of the corresponding Schwinger-Dyson equations for the
unperturbed action is:
\begin{eqnarray}
\langle\zeta^i(w,\bar w)\eta_j(z,\bar z)\rangle_0 &=& 
\frac{\hbar}{2\pi}\delta_j^i\left (-\frac{\dd z}{w-z}+
f_j(w,z)\dd z +g_j(w,\bar z)\dd\bar z\right )\cr
\langle\bar\zeta^i(w,\bar w)\eta_j(z,\bar z)\rangle_0 &=& 
\frac{\hbar}{2\pi}\delta_j^i\left (\frac{\dd \bar z}{\bar w-\bar z}+
\bar f_j(\bar w,\bar z)\dd \bar z +\bar g_j(\bar w,z)\dd z\right)
\end{eqnarray}
where no sum in $j$ is assumed and $f_j, \bar f_j, g_j, \bar g_j$ are
holomorphic in their arguments with domains given by $w, z\in
H_+$. The boundary conditions imply $f_a=f_\mu=\bar f_a=\bar
f_\mu=0$ and
$$g_a(w,\bar z)=-g_\mu(w,\bar z)=\frac{1}{w- \bar z},
\qquad \bar g_a(\bar w, z)=
-\bar g_{\mu}(\bar w,z)=\frac{-1}{\bar w- z}$$

However, if we try to fulfill the boundary conditions for the
components corresponding to the second-class constraints ($A,B,...$
indices) we find a contradiction as $f_A(w, z)$ and $\bar f_A(w, z)$
must extend to entire functions in $w$ and $z$ and besides
$$\bar f_A(w,\bar z) - f_A(w,\bar z) = -\frac{2}{w-\bar z}$$
which is obviously impossible and the propagator does not exist.

At this point one might be tempted to conclude that only the
coisotropic branes make sense at the quantum level. But this is a too
sloppy conclusion since we have only shown that the perturbative
expansion defined by the choice of (\ref{unpgfaccion}) as the
unperturbed part ceases to exist when second-class constraints
appear. The question is whether there is a different definition of the
perturbative expansion leading to a well-defined result in this case.
The situation is reminiscent to the contradiction found by Dirac in
imposing the second-class constraints on the states of the physical
Hilbert space; he proposed to circumvent this difficulty with the help
of the Dirac bracket \cite{Dir}.

{}From now on we shall restrict to branes which satisfy the {\bf strong
regularity condition} (\ref{regularity}) and, in order to make the
presentation simpler, we shall assume that there are no first-class
constraints, i.e. we restrict to the {\bf second-class
branes} of section 2.1.
 
The strategy for solving the problem is the well-known technique of
using our opponent's strength against him. The origin of the
non-existence of the propagator for the second-class coordinates is
that $\det(\Pi^{AB})\not=0$ implies that if $X^A=0$ at the boundary
then $\eta_{A{\bf t}}$ must also vanish. And the propagator cannot satisfy
these two conditions simultaneously. But precisely due to the fact
that $\det(\Pi^{AB})\not=0$ and given that the $\eta_{A}$ fields appear at
most quadratically in the gauge fixed action (\ref{gfaccion}) we can
perform the Gaussian integration over them in order to get an effective
action $S^{eff}$. This action can be used to compute the correlation
functions of observables that do not involve $\eta_A$ fields as it is
our case.

Once the integration has been performed 
there is a splitting of $S^{eff}$
which defines a consistent perturbative expansion. Take $S^{eff}=
S^{eff}_0 +S^{eff}_{pert}$ with
\begin{eqnarray}
S^{eff}_0=\int_D \eta_a\wedge\dd \xi^a 
-\dd{*\eta}_a \lambda^a
+ \omega_{AB}(x)\dd\xi^A\wedge *\dd\lambda^B
- *\dd\gamma^i 
\wedge\dd \beta_i
\end{eqnarray}

The $\beta$, $\gamma$ propagators are as before (see eq.
(\ref{betagamma})), as well as those for $\zeta^a$ and $\eta_a$. In
addition, $S^{eff}_0$ yields well-defined propagators
for the other fields, the only non-zero components being
\begin{eqnarray}
\langle\lambda^A(w,\bar w)\xi^B(z,\bar z)\rangle_0^{eff}
&=&
\frac{i\hbar}{2\pi}\Pi^{AB}(x)\log\frac{|w-z|}{|w-\bar z|}.
\end{eqnarray} 

Now one can expand $S^{eff}_{pert}$ into vertices and define a
perturbative expansion for Green's functions of the form
(\ref{GreenFunctions}). However, from the resulting perturbative
series it seems very hard to find out whether the formula
(\ref{GreenFunctions}) defines an associative product. 
A simpler derivation that gives a positive answer
is given in the next subsection.

\subsubsection{Second-class branes, Kontsevich's formula and Dirac bracket}

Let us take advantage of our opponent's strength in a more profound
sense. Using that $\Pi^{AB}$ is invertible in every point of $C$ and
consequently in a tubular neighborhood of $C$ we can show that the
gauge fixing
\begin{eqnarray}\label{GaugeFixing}
\dd{*\eta_a}=0,\qquad \xi^A=0
\end{eqnarray}
is reachable, at least locally:

$\dd{*\eta_a}=0$ can be obtained by choosing suitably
$\epsilon_a$ in (\ref{symmetryb}). Now, write (\ref{symmetrya}) for
upper-case Latin indices
$$\delta_{\epsilon}\xi^{A}=\Pi^{A{a}}(x+\xi)\epsilon_{{a}}+
\Pi^{AB}(x+\xi)\epsilon_{B}.$$
Since $\Pi^{AB}$ is invertible one can solve for $\epsilon_B$ and get
$\xi^A=0$.
\vskip .2cm
We want to stress that the analog of (\ref{GaugeFixing}) is not an
admissible gauge-fixing in the coisotropic case. For second-class
branes both the Lorentz gauge and (\ref{GaugeFixing}) are admissible
but, as we shall see, the latter makes the perturbative quantization
transparent and is the appropriate approach to the problem.

Let us go back to the BV action (\ref{SBV}), set the indices of the
antighosts $\gamma$ and Lagrange multipliers $\lambda$ upstairs or
downstairs as demanded by (\ref{GaugeFixing}) and take
\begin{eqnarray}\label{Psi}
\Psi= \int_D \gamma^{a}\dd{*\eta}_{a} + \int_D\gamma_A X^A
\end{eqnarray}
where $\gamma_A$ are anticommuting 2-form fields on $D$. 

On the submanifold $\varphi_i^+ =
\frac{\overrightarrow{\delta}\Psi}{\delta \varphi^i}$ we have
\begin{subequations}
\begin{align}
&\beta^{+{a}}=\beta^{+A}=0 \notag \\
&\eta^{+{a}}=*\dd \gamma^{a},\eta^{+A}=0 \notag \\
&X^+_{a}=0,\ X^+_A=-\gamma_A \notag \\
&\gamma^+_{a}=\dd {*\eta}_{a},\ \gamma^{+A}=X^A \notag  
\end{align}
\end{subequations}
\addtocounter{equation}{-1}
And the gauge fixed action with the antifields integrated out reads
now
\begin{eqnarray} \label{Stildegf}
\tilde S_{gf}&=&\int_D \eta_{i}\wedge\dd X^{i} +
\frac{1}{2}\Pi^{ij}(X)\eta_i\wedge\eta_j-*\dd \gamma^{a}\wedge(\dd
\beta_{a}+\partial_{a}\Pi^{kl}(X)\eta_k\beta_l)- \notag\\
&-&\frac{1}{4}{*\dd\gamma}^{a}\wedge *{\dd
\gamma}^{b}\partial_{a}\partial_{b}\Pi^{kl}(X)\beta_k\beta_l-
\lambda^{a}\dd{*\eta_{a}}-\gamma_A\Pi^{Ai}(X)\beta_i
- \lambda_AX^A \notag
\end{eqnarray}
Recall that we are interested in calculating the expectation value of 
functionals depending only on $X$. Hence, integration over 
$\lambda_A$ sets $X^A=0$ and we can
write:
\begin{eqnarray} \label{Stildegf1}
\tilde S_{gf}'&=&\int_D \eta_{a}\wedge\dd X^{a} + 
\frac{1}{2}\Pi^{ij}(X)\eta_i\wedge\eta_j-{*\dd\gamma}^{a}\wedge(\dd \beta_{a}+
\partial_{a}\Pi^{kl}(X)\eta_k\beta_l)- \notag\\
&-&\frac{1}{4}{*\dd\gamma}^{a}\wedge {*\dd\gamma}^{b}\partial_{a}\partial_{b}
\Pi^{kl}(X)\beta_k\beta_l-\lambda^{a}\dd{*\eta}_{a}-
*\gamma_A\Pi^{Ai}(X)\beta_i \notag
\end{eqnarray}
where $\Pi$ is evaluated on $X^A=0$.

Now, integrating over $\gamma_A$ forces
\begin{eqnarray} \label{condition}
&&\Pi^{Ai}\beta_i=0 \Leftrightarrow \beta_A =-
\omega_{AB}\Pi^{B{a}}\beta_{a}
\end{eqnarray}
which is a crucial relation which can be used to get rid of the
components of the fields with upper-case indices and get an effective
action depending only on the lower-case components. Notice that
writing
$$*\dd\gamma^{a}\wedge\partial_{a}\Pi^{kl}(X)\eta_k\beta_l=
*\dd\gamma^{a}\wedge\partial_{a}(\Pi^{{b}{c}}(X)\eta_{b}\beta_{c}
+ \Pi^{Ai}(X)\eta_A\beta_i+\Pi^{{b} A}(X)\eta_{b}\beta_A)$$ and
applying (\ref{condition}) to the second and third terms in
parentheses we obtain the Dirac Poisson structure
(\ref{DiracComponents}) in a beautiful way:
$$*\dd\gamma^{a}\wedge\partial_{a}\Pi^{kl}\eta_k\beta_l=
*\dd\gamma^{a}\wedge\partial_{a}\Pi^{{b}{c}}_{\cal D}\eta_{b}\beta_{c}$$

Doing the same for the term quadratic in $\dd\gamma$ we get:
\begin{eqnarray} \label{SBVeff2}
\tilde S_{gf}''&=&\int_D \eta_{a}\wedge\dd X^{a} + \frac{1}{2}\Pi^{ij}(X)
\eta_i\wedge\eta_j-*\dd \gamma^{a}\wedge(\dd \beta_{a}+\partial_{a}
\Pi_{\cal D}^{bc}(X)\eta_b\beta_c)- \notag\\
&-&\frac{1}{4}{*\dd\gamma}^{a}\wedge *\dd \gamma^{b}\partial_{a}
\partial_{b}\Pi_{\cal D}^{cd}(X)\beta_c\beta_d-\lambda^{a}
\dd{*\eta_{a}}-i\hbar\log\det(\Pi^{AB}(X)) \notag
\end{eqnarray}
where the last term in the action comes from the Jacobian corresponding 
to the delta distribution
$$\delta(\Pi^{Ai}\beta_i)=\delta(\beta_A +
\omega_{AB}\Pi^{B{a}}\beta_{a})\det (\Pi^{BC}(X))$$

The final step is to integrate out the $\eta_A$ fields.  The integral
is Gaussian (due again to the non-degeneracy of $\Pi^{AB}$) and the
determinant coming from it cancels the contribution from the $\delta$
function. Finally,
\begin{eqnarray} \label{SBVeff3}
\tilde S_{gf}^{eff}&=&\int_D \eta_{a}\wedge\dd X^{a} +
\frac{1}{2}\Pi_{\cal D}^{ab}(X)\eta_a\wedge\eta_b-*\dd \gamma^{a}\wedge(\dd
\beta_{a}+\partial_{a}\Pi_{\cal D}^{cd}(X)\eta_c\beta_d)- \notag\\
&-&\frac{1}{4}*{\dd\gamma}^{a}\wedge *\dd
\gamma^{b}\partial_{a}\partial_{b}\Pi_{\cal D}^{cd}(X)
\beta_c\beta_d-\lambda^{a}\dd{*\eta_{a}}
\end{eqnarray}
which is Cattaneo and Felder's gauge-fixed BV action for a
Poisson-Sigma model defined on $D$, with target $(C,\Pi_{\cal D})$
(recall that we set $X^A=0$) and boundary conditions such that $X^{a}$
is free and $\eta_{a}$ vanishes on vectors tangent to $\partial D$. In
other words, we have ended up with the situation studied in
\cite{CatFel99}. Invoking the results therein we can deduce our
announced relation, namely that the perturbative expansion of
$$\left<f(X(0))g(X(1))\delta(X(\infty)-x)\right>,\ f,g\in C^\infty(M)$$
yields Kontsevich's formula for $\Pi_{\cal D}$ applied to the restrictions 
to $C$ of $f$ and $g$. 

In the derivation of this result the second-class boundary conditions
seem to play no role, as they are not used to compute any
propagator. Notice, however, that the gauge fixing (\ref{GaugeFixing})
makes sense only if the fields $\xi^A$ vanish at the boundary before
fixing the gauge. As stressed above the fact that
$\det(\Pi^{AB})\not=0$ is also essential. This ties inextricably the
present result and the use of second-class branes together.

We would like to stress the interesting cancellation of the
determinant coming from the integration of the $\gamma_A$ and
$\beta_A$ fields with that coming from the integration of the $\eta_A$
fields. It would be worth finding out whether there is some underlying
symmetry behind it.

\subsubsection{Quantization with a general strongly regular brane}

Once the quantization of the Poisson-Sigma model with a second-class
brane has been understood, it is straightforward to describe the
procedure for the quantization of the model with an arbitrary {\it
strongly regular brane}\footnote{A brane satisfying the strong
regularity condition.}  defined by both first and second-class
constraints. The appropriate gauge fixing fermion in the general case
is
\begin{eqnarray}\label{generalPsi}
\Psi= \int_D \gamma^{a}\dd{*\eta}_{a} + \gamma^{\mu}\dd{*\eta}_{\mu} +
\gamma_A X^A
\end{eqnarray}

Then, we integrate out the second-class components of the fields
exactly as above and we are left with
\begin{eqnarray} \label{generalSBVeff3}
\tilde S_{gf}^{eff}&=&\int_D \eta_{p}\wedge\dd X^{p} +
\frac{1}{2}\Pi_{\cal D}^{pq}(X)\eta_p\wedge\eta_q-*\dd \gamma^{p}\wedge(\dd
\beta_{p}+\partial_{p}\Pi_{\cal D}^{qr}(X)\eta_q\beta_r)- \notag\\
&-&\frac{1}{4}*{\dd\gamma}^{p}\wedge *\dd
\gamma^{q}\partial_{p}\partial_{q}\Pi_{\cal D}^{rs}(X)
\beta_r\beta_s-\lambda^{p}\dd{*\eta_{p}}
\end{eqnarray}
where the indices now run over $a$ and $\mu$ values and $\Pi_{\cal D}$
is the Dirac bracket of (\ref{DiracCompFirst}). This is Cattaneo and Felder's
gauge-fixed BV action for the Poisson-Sigma model with target given by
local coordinates $(X^a,X^\mu)$, Poisson structure $\Pi_{\cal D}$ and
a coisotropic brane defined by $X^\mu=0$. At this point we can apply
the results of \cite{CatFel03}.

An interesting question is how the choice of a set of second-class
constraints affects the final result. In our case the choice of second
class constraints was made through the gauge fixing fermion,
i.e. a different choice amounts to a change of
gauge-fixing. Since the expectation values of gauge-invariant
observables do not depend on this particular choice, we conclude that
the final result is independent of the choice of second-class
constraints.

The whole derivation parallels that of Dirac's quantization of
constrained systems (\cite{Dir}): one gets rid of the second-class
constraints by defining the appropriate Dirac bracket that can be
quantized with the first class constraints imposed on the states.  It
is nice that this result is obtained in a quantum-field theoretical
context by tuning the boundary conditions of the fields.

\vskip 0.2 cm

{\it Remark:} The need for strong regularity in the quantum case can
be seen, for example, from the fact that we need $\det(\Pi^{AB})\neq
0$ in every point of a tubular neighborhood of $C$ in order to perform
the Gaussian integration over the $\eta_A$ fields.

\section{Conclusions and further work}

We have proven that non-coisotropic branes in the Poisson-Sigma model
are allowed not only classically but also in the quantum
setup. However, the quantization of the model when second-class
constraints are present requires a procedure with differs from the
coisotropic case. In particular, the perturbative expansion must be
redefined, as pointed out in \cite{CalFal05}. After such redefinition
and carrying out the calculations with a suitable gauge-fixing we can
show that the perturbative quantization of the model on the disc with
a second-class brane gives Kontsevich's formula for the Dirac
bracket induced on the brane. In a sense these branes are much simpler
than the coisotropic ones as its quantization always defines an
associative star product and the deformed algebra is defined for all
functions restricted to the brane (no ``quantum'' gauge invariance is
required).

Our final result is an expansion in powers of the Dirac
Poisson tensor $\Pi_{\cal D}$. This fact gives an explanation for the
non-existence of propagator in the perturbative expansion around zero
Poisson structure, which is the one used in the coisotropic case. The
formula (\ref{DiracComponents}) shows that the inverse of $\Pi^{AB}$
enters in the expression of $\Pi_{\cal D}$ and therefore it is not
perturbatively connected to $\Pi=0$.

It is interesting how the boundary conditions of the model have such a
strong influence in the perturbative expansion. It is somehow similar
to the instanton calculations in which one expands around different
classical solutions. Here the (fixed) branes play the role of
instantons determining the perturbative expansion.

In this paper we studied both classical and quantum branes.  In the
classical scenario we saw that {\it weakly regular branes} were
allowed generalizing slightly the results of
\cite{CalFal04}. In this case the phase space of the theory is not a
Poisson manifold since it has special points (defects) in which the
Poisson structure does not exist. The classical theory is then more
efficiently described in terms of a Poisson algebra. Quantization of
these branes, which is not available at the moment, might correspond
to the deformation quantization of a Poisson algebra (rather than a
Poisson manifold), an issue that have some interest on its own. It
would be worth applying these results to the case in which the target
manifold is a Poisson-Lie group and trying to extend the bulk-boundary
duality found in \cite{CalFalGar} to more general boundary conditions.

Another interesting topic is that of the relative positions of
different branes.  In particular one might deform a brane and study
how the Green's functions change. In this way one could try to obtain
the coisotropic brane as a (singular) limit of the second-class one,
which might help to understand the obstructions to associativity and
other technical details found in \cite{CatFel03}.
 
This paper gives a mechanism to carry out the quantization of
second-class submanifolds.  The procedure is somehow indirect and uses
the Poisson-Sigma model as the key ingredient. In the coisotropic case
the reduction has been performed directly at the algebraic level
(\cite{Bor} \cite{CatFel05}), providing a quantum counterpart of the
classical Poisson reduction. It would be very interesting to
investigate this quantum reduction also for non-coisotropic
submanifolds.

\vskip 4mm
\noindent{\bf Acknowledgments:} I. C. thanks G. Felder at the ETH
Zurich and A. S. Cattaneo and M. Zambon at the University of Zurich
for many useful discussions during the preparation of the final
version of this paper.

\end{document}